# Effective Third-Order Nonlinearities in Metallic Refractory Titanium Nitride Thin Films

Nathaniel Kinsey,[1] Akbar Ali Syed,[2] Devon Courtwright,[3] Clayton DeVault,[4]
Carl E. Bonner,[3] Vladimir I. Gavrilenko,[3] Vladimir M. Shalaev,[1] David J. Hagan,[2,5]
Eric W. Van Stryland,[2,5] and Alexandra Boltasseva[1,*]

[1]*School of Electrical and Computer Engineering and Birck Nanotechnology Center, Purdue University, 1205 West State Street, West Lafayette, Indiana, 47907 USA.*
[2]*CREOL, College of Optics and Photonics, University of Central Florida, Orlando, Florida 32816, USA.*
[3]*Center for Materials Research, Norfolk State University, 700 Park Ave, Norfolk, VA, 23504.*
[4]*Department of Physics and Astronomy and Birck Nanotechnology Center, Purdue University, West Lafayette, Indiana, 47907, USA.*
[5]*Department of Physics, University of Central Florida, Orlando, Florida 32816, USA.*
*aeb@purdue.edu*

**Abstract:** Nanophotonic devices offer an unprecedented ability to concentrate light into small volumes which can greatly increase nonlinear effects. However, traditional plasmonic materials suffer from low damage thresholds and are not compatible with standard semiconductor technology. Here we study the nonlinear optical properties in the novel refractory plasmonic material titanium nitride using the Z-scan method at 1550 nm and 780 nm. We compare the extracted nonlinear parameters for TiN with previous works on noble metals and note a similarly large nonlinear optical response. However, TiN films have been shown to exhibit a damage threshold up to an order of magnitude higher than gold films of a similar thickness, while also being robust, cost-efficient, bio- and CMOS-compatible. Together, these properties make TiN a promising material for metal-based nonlinear optics.

## 1. Introduction

Among the many materials used in nonlinear optics, traditional metals have long been known to exhibit large nonlinear coefficients [1] and offer the potential for significant field enhancement when nanostructured [2-4]. Consequently, the role of metals in nonlinear optics can be divided into two regimes: 1) when the metal itself serves as the nonlinear medium and 2) when metal serves as a supplementary element for a nonlinear system. Towards the first point, many proof-of-concept demonstrations of metallic nonlinear devices such as frequency conversion [1, 5-9], ultrafast dynamic switching [10, 11], high sensitivity biological detectors [12-14], and enhanced spectroscopy [15] have been reported. However, the example devices mentioned above, which rely on the nonlinearities in metals, have not seen widespread adoption which may be due in part to their low efficiency and propensity for deformation under the intense fields required for nonlinear optics [16, 17]. Subsequently, efforts have been directed towards point two where metallic components are supplementary to another, more efficient, nonlinear medium and are designed, for instance, to enhance or concentrate the electric field [18]. However, even in this case, the capability of such structures to withstand the intense fields generated by confinement is limited. Thus, there is a need to look for better materials which are not only plasmonic, but possess the ability to withstand high intensities, and to understand the inherent nonlinearities present in such materials.

Recently, TiN has been suggested as a refractory metal (melting point > 2900°C) with plasmonic properties similar to gold [19]. In addition, TiN has tunable optical properties, is chemically stable, can be grown epitaxially on magnesium oxide, c-sapphire, and silicon, and is bio- and CMOS-compatible, all in stark contrast to the noble metals [19, 20]. In fact, TiN-based metasurfaces have been experimentally demonstrated to withstand temperatures and optical intensities greater than gold structures, making them potentially interesting for applications in nonlinear optics [21]. However, the inherent nonlinearities of this important material have yet to be investigated, although some studies have been conducted on weakly plasmonic nanoparticle matrices [22, 23]. These studies do not provide information upon the inherent nonlinearities in the metal as it is known that nanostructured samples can exhibit altered nonlinearities due to geometric parameters (for example, plasmon resonances) [23, 24]. Additionally, the study of S. Divya *et al* used nanosecond pulses where cumulative thermal effects (i.e. increased lattice temperature) can significantly contribute to the observed nonlinearities. Here we extract the ultrafast nonlinearities using femtosecond pulses on thin films of TiN, enabling characterization of the underlying inherent material nonlinearities which describe the response of the material in the absence of external parameters such as nanostructuring (e.g. surface plasmon resonance) or enhancement (e.g. field confinement). Using the dual-arm Z-scan technique at both 1550 nm and 780 nm, we find nonlinearities in TiN films which are similar to the large nonlinearities found in traditional metals.

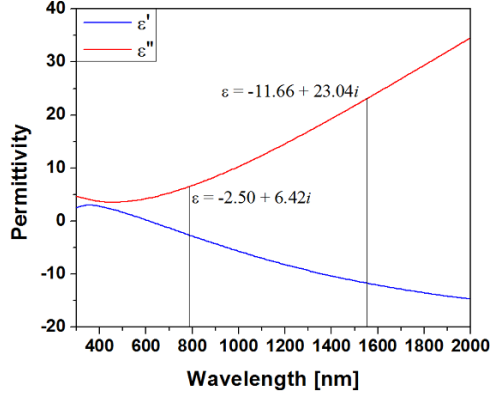

Fig. 1. Linear optical spectra of TiN deposited at 350°C on silica glass as derived from spectroscopic ellipsometry measurements.

## 2. Results

A 52 nm thick TiN film deposited on fused silica at 350°C was investigated in this work. The linear optical functions of the TiN films, shown in Fig. 1, were obtained using spectroscopic ellipsometry and the model as described in Eq. (3) (see the Appendix). The TiN sample is found to have a permittivity of $\varepsilon = -2.50 + i6.42$ ($\tilde{n}_o = 1.48 + i\,2.17$) at 780 nm and $\varepsilon = -11.66 + i\,23.04$ ($\tilde{n}_o = 2.66 + i\,4.33$) at 1550 nm. The nonlinear optical properties were investigated using a dual-arm Z-scan technique (see Appendix for detailed description) [25, 26]. The total complex refractive index including third-order nonlinearities can be written as $\tilde{n} = \tilde{n}_o + \tilde{n}_2 I$ where $\tilde{n}_o = n'_o + i n''_o$ is the complex linear refractive index, $\tilde{n}_2 = n'_2 + i n''_2$ is the complex nonlinear refractive index, and $I$ is the input light intensity. The measurable quantities for the nonlinear refractive index and nonlinear absorption are usually written as $n(I) = n_0 + n_2 I$ and $\alpha(I) = \alpha_0 + \alpha_2 I$ where $dI/dz = -\alpha(I)I$, which are related to $\tilde{n}_2$ by $n'_2 = n_2$ and $n''_2 = (\lambda/4\pi)\alpha_2$. Following the procedure of del Corso and Solisthe, the real and imaginary portions of the third-order susceptibility in SI units are given by [27]:

$$\text{Re}\{\chi^{(3)}\} = \frac{4}{3} n'_o \varepsilon_o c \left[ n'_o n_2 - n''_o \alpha_2 \frac{\lambda}{4\pi} \right] \quad (1)$$

$$\text{Im}\{\chi^{(3)}\} = \frac{4}{3} n'_o \varepsilon_o c \left[ n'_o \alpha_2 \frac{\lambda}{4\pi} + n''_o n_2 \right] \quad (2)$$

where $\varepsilon_o$ is the free space permittivity, $c$ is the speed of light, $\lambda$ is the wavelength, and other parameters are as defined above. We note here that the nonlinear refraction and absorption depend on both the real and imaginary parts of the susceptibility (see Appendix) [28, 29]. The common approximations that $\text{Re}\{\tilde{\chi}^{(3)}\} \propto n_2$ and $\text{Im}\{\tilde{\chi}^{(3)}\} \propto \alpha_2$ do not apply in metal films where the imaginary part of the index is large. It is also important to note that the incident intensity values should be corrected for the reflectance of the multilayer system, such that $I_{eff} = I_o(1 - R)$ [29]. For the TiN on fused silica, $R = 0.412$ at 780 nm and $R = 0.587$ at 1550 nm which are determined from the linear optical properties using the transfer matrix method for thin films [30].

Additionally, the measurements were completed for excitation pulse widths of 95 fs at 1550 nm and 220 fs at 780 nm and the extracted nonlinear parameters may vary for pulse widths different from these values. Specifically, thermal nonlinearities within the pulse envelope become important as the pulse width nears or exceeds a critical value given by $t_p \geq \rho_o C/(dn/dT)\alpha_o$ where $n_2$ is the nonlinear refractive index, $\rho_o$ is the density, $C$ is the heat capacity, $dn/dT$ is the temperature dependent refractive index change (i.e. cumulative thermal nonlinearity), and $\alpha_o$ is the absorption coefficient [31]. For TiN and the values $\rho_o C = 3.13 \times 10^6$ [J/Km$^3$] [32, 33], $dn/dT = 6 \times 10^{-4}$ [K$^{-1}$] [34], $n_2$ values as shown below, and $\alpha_o = 3.5 \times 10^5$ [cm$^{-1}$], we find a critical pulse width of ~500 fs. Thus, even with the current excitation parameters (95 fs and 220 fs), thermal nonlinearities within the pulse envelope may begin to play a role in the measurement, and are likely to result in modified values of the extracted nonlinear properties for pulse widths longer than those used here.

The open and closed aperture Z-scan results at 1550 nm are shown in Fig. 2(a),(b) for several incident intensities ranging from 24 to 141 [GW/cm$^2$] (27 - 155 nJ/pulse) as calculated for a Gaussian pulse by $I_o = 2E_{pulse}/\pi^{3/2}w_o^2\tau$ where $E_{pulse}$ is the pulse energy, $w_o$ is the beam waist, and $\tau$ is the 1/e pulse width given by $\tau = t_{FWHM}/2\sqrt{\ln(2)}$ [26]. Each scan was completed twice and the results were averaged to further reduce any error due to beam instability.

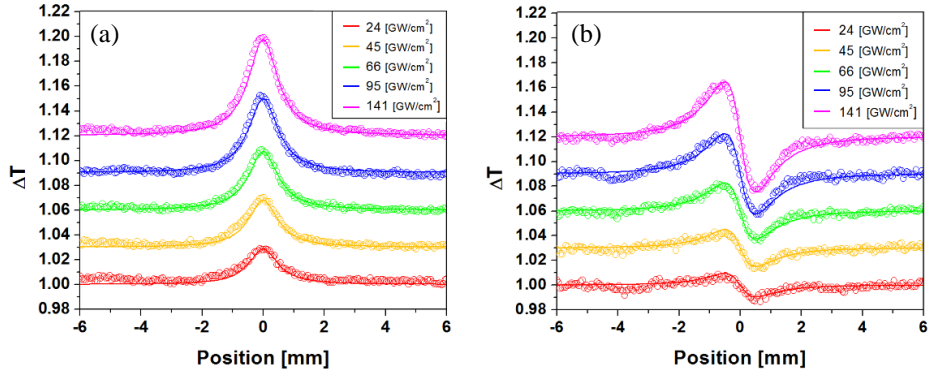

Fig. 2. Compilation of the a) open aperture and b) closed aperture Z-scan curves for several different intensities at 1550 nm. Experimental results are shown with symbols and the fitted curves are depicted with a solid line.

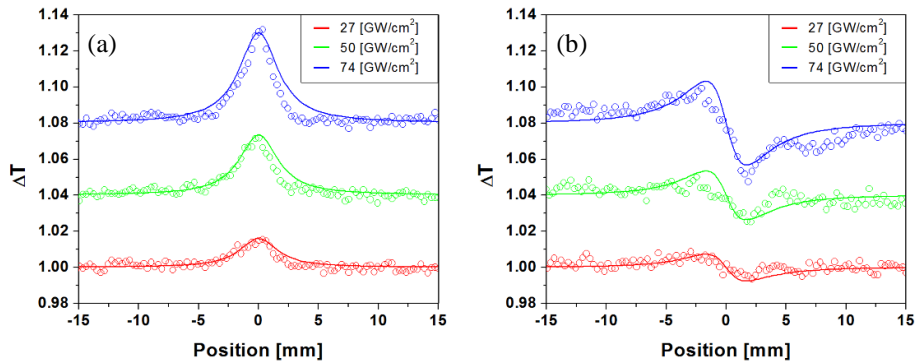

Fig. 3. Compilation of the a) open aperture and b) closed aperture Z-scan curves for several different intensities at 780 nm. Experimental results are shown with symbols and the fitted curves are depicted with a solid line.

The open aperture Z-scan shows saturable absorption described by $\alpha(I) = \alpha_o/(1 + I/I_{sat})$ [35] giving a fitted saturation intensity of $I_{sat} = 530$ [GW/cm$^2$]. Expanding to the first order with $\alpha(I) \approx \alpha_o - (\alpha_o/I_{sat})I = \alpha_o + \alpha_2 I$, an average value of $\alpha_2 = -6.6 \times 10^{-9}$ [m/W] is obtained. Likewise, fitting the closed aperture experimental data, an $n_2 = -3.7 \times 10^{-15}$ [m$^2$/W] is extracted. Using Eq. (1), (2), the total complex third-order susceptibility is found to be $\tilde{\chi}^{(3)}_{eff} = -5.9 \times 10^{-17} - i\, 1.7 \times 10^{-16}$ [m$^2$/V$^2$] or $\tilde{\chi}^{(3)}_{eff} = -4.2 \times 10^{-9} - i\, 1.2 \times 10^{-8}$ [esu].

Likewise under an excitation wavelength of 780 nm the open and closed aperture results are shown in Fig. 3(a),(b) for several intensities. Using the same fitting procedure, values for the nonlinear coefficients were found to be $I_{sat} = 510$ [GW/cm$^2$] resulting in $\alpha_2 = -6.8 \times 10^{-9}$ [m/W] and $n_2 = -1.3 \times 10^{-15}$ [m$^2$/W]. These values result in complex third-order susceptibility of $\tilde{\chi}^{(3)}_{eff} = -5.3 \times 10^{-18} - i\, 1.8 \times 10^{-17}$ [m$^2$/V$^2$] or $\tilde{\chi}^{(3)}_{eff} = -3.8 \times 10^{-10} - i\, 1.3 \times 10^{-9}$ [esu].

## 3. Discussion

The results of our experiments have been summarized in Table 1 along with several other relevant works studying the nonlinear properties of thin metal films and a recent result investigating TiN nanoparticles [22]. We note here the description of the nonlinearity as an effective $\tilde{\chi}^{(3)}$, denoted $\tilde{\chi}^{(3)}_{eff}$. This distinction is made due to the multitude of processes which can contribute to the observed signal in metals and our TiN films such as population rearrangement, band filling, or bandgap renormalization which are not intrinsically third-order processes. However, these processes can be modeled through the complex nonlinear refractive index as has been done for previous metal films, although other methods can potentially be used [36, 37].

In addition, we note that the data available in the literature contains some stark differences to our measurements such that a direct and quantitative comparison is difficult due to varying wavelengths, differing methods of characterization (optical Kerr effect or Z-scan), different pulse widths, and different film thicknesses. First, variations in the wavelength can certainly lead to an altered nonlinear response (as is shown in the TiN film). One example for this variance in films can be the presence of resonant features which can significantly increase the nonlinearities (for instance, the d-sp orbital transition in gold occurring near 500 nm) [38]. Consequently, one may expect that moving to a longer wavelength in gold films (i.e. off resonance) would result in a decrease of the nonlinear response. Secondly, due to the difficulty in completing the closed aperture Z-scan analysis on metal films, some data from literature were obtained using an optical Kerr effect measurement [29]. This measurement, in general, deduces a different tensor value of $\tilde{\chi}^{(3)}$ which need not relate to the value measured with Z-scan. Also, the results from the literature use a significantly longer pulse width than that used here. It is well known for dielectric materials that a longer pulse width can drastically increase the nonlinear response through the incorporation of additional, slower effects such as electrostriction, thermal heating etc. [31] and has also been shown to produce a similar dependence in metal films due to thermal smearing d-band electrons [37]. Finally, as noted by E. Xenogiannopoulou *et al*, thinner films (of a few nm's) can show an enhanced nonlinear response, roughly a factor of 4 to 5 increase when the thickness is decreased from ~50 nm to ~5 nm [29]. Therefore, the nonlinear responses in the thin silver and gold films reported in Table 1 may be increased due to their small thickness.

Despite these factors, we note that the nonlinearities in TiN films are similar in magnitude to other standard metal films. Additionally, it has been shown that TiN can withstand a significant intensity before damage occurs, owing to its properties as a refractory metal. In this previous work, a damage threshold of ~5 [GW/cm$^2$] (0.2 [J/cm$^2$] for 40 ps pulses at 2 Hz and λ = 532 nm) was found [39]. For comparison, gold films are reported to have a damage threshold of $I_o$ ~ 400 [MW/cm$^2$] (14 [mJ/cm$^2$] for 35 ps pulses at 10 Hz and λ = 532 nm), which is one order of magnitude less than that of TiN films [28, 29].

Additionally, we note that TiN can be grown epitaxially on silicon, c-sapphire, and MgO, enabling high-quality ultra-thin films down to 2 nm which can increase the nonlinear response of the material [41].

While thinner films are likely to have a lower damage threshold, such a TiN film may also increase the nonlinearity, as has been documented with other metallic films (although this effect may be different for femtosecond pulses). Also, due to the aforementioned d-sp transition in gold, open aperture Z-scans of gold films observe two-photon absorption in the range of 532 - 1064 nm. However, we note that TiN exhibits saturable absorption even as low as 780 nm. This is due to the lack of any resonant absorptive term in the permittivity until shorter wavelengths less than 400 nm. This situation is similar to the case of silver, which also exhibits saturable absorption even as high as 532 nm, and may be useful for applications towards TiN-based intensity selective mirrors used in mode-locked lasers where both high reflectivity and saturable absorption can be achieved in a single thin film.

Table 1. Comparison of the third-order susceptibilities of thin metal films. Some results used the simplified relations for $\tilde{\chi}_{eff}^{(3)}$ (marked with *). The results for silver film were recalculated using the full complex relationship of $\tilde{\chi}_{eff}^{(3)}$ (marked with **) using the refractive index of silver at 532 nm as found from literature (since the value was not provided in the paper) [40].

| Material | λ [nm] | Pulse-Width | $α_o$ [cm$^{-1}$] | Re{$\tilde{\chi}_{eff}^{(3)}$} [esu] | Im{$\tilde{\chi}_{eff}^{(3)}$} [esu] |
|---|---|---|---|---|---|
| 52 nm TiN film on Fused Silica | 1550 | 95 fs | $3.5 \times 10^5$ | $-4.2 \times 10^{-9}$ | $-1.2 \times 10^{-8}$ |
| | 780 | 220 fs | $3.5 \times 10^5$ | $-3.8 \times 10^{-10}$ | $-1.3 \times 10^{-9}$ |
| 55 nm TiN/PVA nanoparticle matrix [23] | 532 | 7 ns | $5.8 \times 10^5$ | $-1.9 \times 10^{-11}$ | $5.0 \times 10^{-11}$ |
| 5 nm Au film [28] | 532 | 30 ps | $5.7 \times 10^5$ | - | $8.6 \times 10^{-8}$ * |
| 52 nm Au film [29] | 532 | 35 ps | $3.3 \times 10^5$ | $7.0 \times 10^{-10*}$ | $4.0 \times 10^{-9}$ * |
| 8 nm Ag film [36] | 532 | 10 ns | $2.9 \times 10^5$ | $-6.4 \times 10^{-8**}$ | $2.6 \times 10^{-7**}$ |

## 4. Conclusion

In this work, we have investigated the nonlinear refraction and absorption of the novel refractory metal TiN by the dual-arm Z-scan method at the technologically important wavelengths of 1550 nm and 780 nm. It is found that the effective third-order nonlinear optical susceptibility values are similar to other traditional metal films as well as TiN nanoparticles. However, unlike gold films, TiN is shown to have saturable absorption behavior up to 780 nm with saturation intensities of ~500 [GW/cm$^2$]. Additionally, previous demonstrations illustrate that TiN films can withstand intensities of ~5 [GW/cm$^2$] (40 ps pulses), an order of magnitude larger than is reported in gold films for similar excitation conditions. Collectively, these properties make TiN a promising material for practical applications using metal-based nonlinear devices.

### Acknowledgments

The authors appreciate helpful discussions and manuscript revisions from Prof. M. Noginov (Norfolk State University), Prof. M. Ferrera (Purdue University/Heriot-Watt University), and Dr. A. Lagutchev (Purdue University). This work is supported by ARO grant 57981-PHW911NF-11-1-0359, ARO MURI grant 56154-PH-MUR (W911NF-09-1-0539), NSF PREM DRM-1205457, NSF NCN EEC-0228390, AFOSR FA9550-09-1-0456.

### References

1. N. Bloembergen, W. Burns, and M. Matsuoka, "Reflected third harmonic generated by picosecond laser pulses," Opt. Commun. **1** 1-4 (1969).
2. M. Kauranen, and A. V. Zayats, "Nonlinear Plasmonics," Nature Photon. **6** 737-748 (2012).
3. M. I. Stockman, "Nanoplasmonics: past, present and glimpse into future," Opt. Express **19**(22), 22029-22106 (2011).
4. S. A. Maier, *Plasmonics: Fundamentals and Applications* (Springer, 2007).

5. J. E. Sipe, V. C. Y. So, M. Fukui, and G. I. Stegeman, "Analysis of second-harmonic generation at metal surfaces," Phys. Rev. B **21**(10), 4389-4403 (2980).
6. M. W. Klein, C. Enkrich, M. Wegener, and S. Linden, "Second-Harmonic Generation From Magnetic Metamaterials," Science **313** 502-504 (2006).
7. I. V. Shadrivov, A. A. Zharov, and Y. S. Kivshar, "Second-harmonic generation in nonlinear left-handed metamaterials," J. Opt. Soc. Am. B **23** 529-534 (2006).
8. M. D. McMahon, R. Lopez, R. F. H. Jr., E. A. Ray, and P. H. Bunton, "Second-harmonic generation from arrays of symmetric gold nanoparticles," Phys. Rev. B **73** 041401-041405 (2006).
9. I. Y. Park, S. Kim, J. Choi, D. H. Lee, Y. J. Kim, M. F. Kling, M. I. Stockman, and S. W. Kim, "Plasmonic generation of ultrashort extreme-ultraviolet light pulses," Nature Photon. **5** 677-681 (2011).
10. K. MacDonald, Z. Sámson, M. Stockman, and N. Zheludev, "Ultrafast active plasmonics," Nature Photon. **3**(1), 55-58 (2008).
11. A. V. Krasavin, T. P. Vo, W. Dickson, P. M. Bolger, and A. V. Zayats, "All-plasmonic modulation via stimulated emission of copropagating surface plasmon polaritons on a substrate with gain," Nano Lett. **11**(6), 2231-2235 (2011).
12. B. Sharma, R. R. Frontiera, A. I. Henry, E. Ringe, and R. P. van Duyne, "SERS: Materials, applications, and the future," Mater. Today **15** 16-25 (2012).
13. J. N. Anker, W. P. Hall, O. Lyandres, N. C. Shah, J. Shao, and R. P. van Duyne, "Biosensing with plasmonic nanosensors," Nat. Mater. **7** 442-453 (2008).
14. B. Luk'yanchuk, N. I. Zheludev, S. A. Maier, N. J. Halas, P. Nordlander, H. Giessen, and C. T. Chong, "The Fano resonance in plasmonic nanostructures and metamaterials," Nat. Mater. **9** 707-715 (2010).
15. J. R. Lakowicz, "Plasmonics in biology and plasmon-controlled flourescence," Plasmonics **1** 5-33 (2006).
16. J. B. Khurgin, and G. Sun, "Plasmonic enhancement of the third order nonlinear optical phenomena: Figures of Merit," Opt. Express **21**(22), 27460-27480 (2013).
17. U. Guler, A. Boltasseva, and V. M. Shalaev, "Refractory Plasmonics," Science **344** 263-264 (2014).
18. J. Lee, M. Tymchenko, C. Argyropoulos, P. Y. Chen, F. Lu, F. Demmerle, G. Boehm, M. C. Amann, A. Alu, and M. A. Belkin, "Giant nonlinear response from plasmonic metasurfaces coupled to intersubband transitions," Nature **511** 65-69 (2014).
19. G. V. Naik, V. M. Shalaev, and A. Boltasseva, "Alternative Plasmonic Materials: Beyond Gold and Silver," Adv. Mater. **25**(24), 3264-3294 (2013).
20. D. Starosvetsky, and I. Gotman, "Corrosion behavior of titanium nitride coated Ni-Ti shape memory surgical alloy," Biomaterials **22**(13), 1853-1859 (2001).
21. W. Li, U. Guler, N. Kinsey, G. V. Naik, A. Boltasseva, J. Guan, and V. M. Shalaev, "Refractory Plasmonics with Titanium Nitride: Broadband Metamaterial Absorber," Adv. Mater. **26**(47), 7959-7965 (2014).
22. S. Divya, V. Nampoori, P. Radhakrishnan, and A. Mujeeb, "Evaluation of nonlinear optical parameters of TiN/PVA nanocomposite - A comparison between semi-empirical relation and Z-scan results," Curr. Appl. Phys. **14** 93-95 (2014).
23. S. Divya, V. P. N. Nampoori, P. Radhakrishnan, and A. Mujeeb, "Origin of optical non-linear response in TiN owing to excitation dynamics of surface plasmon resonance electronic oscillations," Laser Phys. Lett. **11**(8), 1-7 (2014).
24. K. Fukumi, A. Chayahara, K. Kadono, T. Sakaguchi, Y. Horino, M. Miya, K. Fujii, J. Hayakawa, and M. Satou, "Gold nanoparticles ion implanted in glass with enhanced nonlinear optical properties," J. Appl. Opt. **75** 3075 (1994).
25. M. Sheik-Bahae, A. A. Said, T.-H. Wei, D. J. Hagan, and E. W. van Stryland, "Sensitive measurement of optical nonlinearities using a single beam," IEEE J. Quant. Electron. **26** 760-769 (1990).
26. M. R. Ferdinandus, M. Reichert, T. R. Ensley, H. Hu, D. A. Fishman, S. Webster, D. J. Hagan, and E. W. van Stryland, "Dual-arm Z-scan technique to extract dilute solute nonlinearities from solution measurements," Opt. Mater. Express **2**(12), 1779-1790 (2012).
27. R. del Coso, and J. Solis, "Relation between nonlinear refractive index and third-order susceptibility in absorbing media," J. Opt. Soc. Am. B **21**(3), 640-644 (2004).
28. D. D. Smith, Y. Yoon, R. W. Boyd, J. K. Campbell, L. A. Baker, R. M. Crooks, and M. George, "Z-Scan Measurement of the Nonlinear Absorption of a Thin Gold Film," J. Appl. Opt. **86** 6200 (1999).
29. E. Xenogiannopoulou, and P. Aloukos, "Third-order nonlinear optical properties of thin sputtered gold films," Opt. Commun. **275** 217-222 (2007).
30. S. Ishii, U. K. Chettiar, X. Ni, and A. V. Kildishev, "PhotonicsRT: Wave propagation in multilayer structures," (2014).
31. R. Boyd, *Nonlinear Optics* (Elsevier, Burlington, MA, 2008).
32. D. Lide, *CRC Handbook of Chemistry and Physics* (CRC Press, Boca Raton, 2005).
33. B. F. Naylor, "High-temperature heat contents of Titanium Carbide and Titanium Nitride," Journal of the American Chemical Society **68**(3), 370-371 (1946).
34. S. T. Sundari, R. Ramaseshan, F. Jose, S. Dash, and A. K. Tyagi, "Investigation of temperature dependent dielectric constant of a sputtered TiN thin film by spectroscopic ellipsometry," J. Appl. Opt. **115** 00335161-00335166 (2014).
35. R. L. Sutherland, D. G. Mclean, and S. Kirkpatrick, *Handbook of Nonlinear Optics* (Marcel Dekker, New York, 2003).
36. G. Yang, D. Guan, W. Wang, W. Wu, and Z. Chen, "The inherent optical nonlinearities of thin silver films," Opt. Mater. **25** 439-443 (2004).
37. N. Rotenberg, A. D. Bristow, M. Pfeiffer, M. Betz, and H. M. van Driel, "Nonlinear absorption in Au films: Role of thermal effects," Phys. Rev. B **75**(15), 155426-155430 (2007).
38. M. M. Alvarez, J. T. Khoury, T. G. Schaaff, M. N. Shafigullin, I. Vezmar, and R. L. Whetten, "Optical Absorption Spectra of Nanocrystal Gold Molecules," J. Phys. Chem. B **101**(19), 3706-3712 (1997).
39. B. Gakovic, M. Trtica, D. Batani, P. Panjan, and D. Vasiljevic-Radovic, "Surface modification of titanium nitride film by a picosecond Nd:YAG laser," J. Opt. A-Pure Appl. Op. **9** S76-S80 (2007).
40. X. Ni, Z. Liu, and A. V. Kildishev, "PhotonicsDB: Optical Constants," (2010).
41. G. V. Naik, B. Saha, J. Liu, S. M. Saber, E. A. Stach, J. M. K. Irudayaraj, T. D. Sands, V. M. Shalaev, and A. Boltasseva, "Epitaxial superlattices with titanium nitride as a plasmonic component for optical hyperbolic metamaterials," Proc. Natl. Academ. Sci. USA **111**(21), 9546-7551 (2014).

**Appendix**

*Fabrication*

The TiN films were fabricated using reactive magnetron sputtering (PVD Systems Inc.) similar to the method described in reference [19]. A titanium target was sputtered into a 60% nitrogen 40% argon environment at 5 mT. The substrate was heated during the deposition to 350°C and the resulting TiN films on fused silica form a polycrystalline structure. In addition, control of the substrate temperature enables the modification of the carrier concentration within the film. Higher temperatures allow more carriers (i.e. cross-over permittivity ~ 500 nm at 800°C versus ~600 nm at 350°C. The linear optical properties of the TiN films were measured using variable angle spectroscopic ellipsometry (J. A. Wollam Co.) at two angles of 50° and 70°. The ellipsometry data were fitted using a Drude+Lorentz model encompassing three oscillators as follows[1]:

$$\varepsilon(\omega) = \varepsilon_\infty - \frac{f_{Drude}\omega_p^2}{\omega^2 + i\Gamma_{Drude}\omega} + \sum_{m=1}^{2} \frac{f_m \omega_m^2}{\omega_m^2 - \omega^2 - i\Gamma_m \omega} \qquad (3)$$

where $\varepsilon_\infty$ is the permittivity at high frequency, $\omega_p$ is the unscreened plasma frequency, $f_m$ and $f_{Drude}$ are the strength of the oscillators, $\omega_m$ is the resonant frequency corresponding to the Lorentz oscillator, and $\Gamma_m$ and $\Gamma_{Drude}$ are the damping of the oscillators. The first term captures the Drude-like metallic response while the other two Lorentz terms capture the absorption peaks.

*Experimental Characterization*

A Ti: Sapphire laser system (Clark-MXR, CPA 2110) at 780 nm with 1 mJ energy/pulse, 150 fs (FWHM) pulse width, and 1 kHz repetition rate was used to pump the optical parametric amplifier (Light conversion, TOPAS-C). The output of TOPAS-C was tuned to 1550 nm which is used for our dual arm Z-Scan measurements [26]. The input beam was sent through the combination of half wave plate and polarizer for fine tuning of the energy then spatial filtered to obtain a Gaussian beam. To monitor laser fluctuations, a small fraction of the laser beam (approximately 10%) was deflected and used as a reference. The remaining 90% was evenly divided into two beams by using a 50/50 beam splitter sent through the Dual-Arm (DA) Z-Scan. We have reported DA Z-Scan measurements for solutions by keeping the solution in one arm and solvent in another arm [26]. We followed the same procedure in the present measurements by replacing the solution with the TiN thin film and the solvent with the bare substrate. The DA Z-scan essentially cancels correlated noise between the two arms (e.g., pulse width, pulse energy and beam pointing) to greatly increase the signal–to-noise ratio. To implement this technique the system which is constructed with two identically Z-scan arms, is first calibrated by placing identical fused silica samples in each arm and adjusting the energy and sample positions to get a null differential Z-scan signal, i.e. Z-scan signals subtracted. Once calibrated we replace the fused silica in the two arms with the TiN thin film and the bare substrate respectively. The closed aperture (CA) DA Z-scan profile of TiN is obtained by subtracting the CA signal of the bare substrate from that of the TiN thin film. Similarly, the open aperture DA Z-scan of TiN was simultaneously measured as described in [26]. The pulse width at 1550 nm and 780 nm was determined from the closed aperture Z-Scan of fused silica. The beam waist at the focus was calculated by performing the open aperture Z-Scan of GaAs and ZnSe which shows 2PA at 1550 nm and at 780 nm, respectively (the FWHM of the open aperture scan is equal to $2z_o$).

*Relation of Complex Susceptibility to Measurable Quantities*

Traditionally, only the real portion of the linear refractive index is used during the calculation of the susceptibility. While this simplification is acceptable in the cases of low-loss dielectrics where $\tilde{n}_o' \gg \tilde{n}_o''$, it cannot be used for metals [28]. Due to the complex nature of the refractive index, coupling between the real (imaginary) nonlinear index and imaginary (real) susceptibility arises. The general relation between the complex third-order susceptibility and the nonlinear refraction is shown in Eq. (4) as derived from reference [31] with $I = 2\varepsilon_o n'_o c |E(\omega)|^2$ following the procedure of [27]. Here we adopt the definitions of $\tilde{n}_2$ and $\tilde{\chi}^{(3)}$ as presented in reference [31], although other definitions are also used in literature.

$$\tilde{\chi}^{(3)} = \frac{4}{3}\varepsilon_o c n'_o \tilde{n}_o \tilde{n}_2 \tag{4}$$

The real and imaginary portions of the susceptibility in SI units are then given by:

$$\text{Re}\{\chi^{(3)}\} = \frac{4}{3} n'_o \varepsilon_o c \left[n'_o n_2 - n''_o \alpha_2 \frac{\lambda}{4\pi}\right] \tag{5}$$

$$\text{Im}\{\chi^{(3)}\} = \frac{4}{3} n'_o \varepsilon_o c \left[n'_o \alpha_2 \frac{\lambda}{4\pi} + n''_o n_2\right] \tag{6}$$

Many works in literature also use the electrostatic unit system where the third-order susceptibility is related to SI units by $\tilde{\chi}^{(3)}[SI] = 1.4 \times 10^{-8} \tilde{\chi}^{(3)}[esu]$ [31]. If the losses in the material are low then we can clearly see that the formulas reduce to the typical form (within a scaling factor that depends upon the initial definitions) as presented in other works [25]:

$$\text{Re}\{\chi^{(3)}\} = \frac{4}{3} {n'_o}^2 \varepsilon_o c n_2 \tag{7}$$

$$\text{Im}\{\chi^{(3)}\} = \frac{4}{3} {n'_o}^2 \varepsilon_o c \alpha_2 \frac{\lambda}{4\pi} \tag{8}$$

However, as we have mentioned, these simplified formulas are not a fully accurate description of the third-order susceptibility for lossy films, and Eq. (5),(6) should be used in general.